\documentclass[a4paper,11pt]{article}
\usepackage{pos}
\renewcommand{\logo}{\vbox{\hfill{\tt BU-HEPP-24-06}}\vspace{-2cm}}
\renewcommand{\posurl}{}
\title{KKMCee: Multi-photon MC for lepton and quark pair production at lepton colliders}
\ShortTitle{KKMCee for lepton and quark pair production at lepton colliders}
\author*[a]{Scott A. Yost}
\author[1]{Stanisław Jadach}
\author[b]{B.F.L. Ward}
\author[c]{Zbigniew W{\c a}s}
\author[d]{Andrzej Si{\'o}dmok}
\affiliation[a]{Physics Department, The Citadel,\\
  171 Moultrie St., Charleston, SC 29409, U.S.A.}
\affiliation[b]{Physics Department, Baylor University,\\
1311 S 5th St., Waco, TX 76706, U.S.A.}
\affiliation[c]{Institute of Nuclear Physics Polish Academy of Science,\\
ul. Radzikowskiego 152 31-342 Krak{\'o}w, Poland}
\affiliation[d]{Department of Application of Computational Methods, Jagiellonian University,\\
ul. Łojasiewicza 11, 30-348 Krak{\'o}w, Poland}
\emailAdd{scott.yost@citadel.edu}
\emailAdd{bfl\_ward@baylor.edu}
\emailAdd{zbigniew.was@ifj.edu.pl}
\emailAdd{andrzej.siodmok@uj.edu.pl}
\abstract{We present an overview of the Monte Carlo event generator KKMCee for lepton and quark pair production in the high energy electron-positron annihilation process. We note that it is still the most sophisticated event generator for such processes. Its entire source code has been rewritten in C++. We have verified that the new version, KKMCee 5.00, reproduces the benchmarks of the older code in FORTRAN 77. We discuss a number of improvements in the MC algorithm and its interfaces.}
 \note{Deceased}
\FullConference{42nd International Conference on High Energy Physics (ICHEP2024)\\
18-24 July 2024\\
Prague, Czech Republic\\}
\begin{document}
\maketitle
This note summarizes the updates in the MC program KKMCee 5.00~\cite{KKMCee:2023} for precision electroweak phenomenology in $e^+e^-$ collisions.  KKMCee features exponentiated multiple photon emission, $e^+e^- \rightarrow Z/\gamma^* \rightarrow f{\overline f} + n\gamma$
including exact $O(\alpha)$ and $O(\alpha^2 L)$ hard-photon residuals, where $L$ is the appropriate large logarithm, for initial-state radiation (ISR), final-state radiation (FSR) and initial-final interference (IFI). KKMCee uses an amplitude-level implementation of soft photon exponentiation, ``CEEX,''~\cite{Jadach:1998jb,Jadach:2000ir} to be distinguished from ``EEX,'' the cross-section level version known also as YFS exponentiation~\cite{YFS:1961} which is also supported for cross-checks.\footnote{EEX mode does not support IFI, but adds $O(\alpha^3 L^3)$ leading logarithm corrections that are not implemented in CEEX.} KKMCee supports collision energies up to 1 TeV and had a LEP2 precision tag of $0.2\%$.~\cite{Jadach:1999vf} 

Electroweak corrections are included via a DIZET module~\cite{Bardin:1989tq,Arbuzov:2020coe} and tau decays are implemented using a TAUOLA module.~\cite{Jadach:1993hs,Chrzaszcz:2016fte} There is now a HEPMC3 event record,~\cite{Buckley:2019xhk} which is instrumental for interfacing to the latest version of PHOTOS~\cite{Barberio:1990ms,Davidson:2010ew} and facilitates interfacing to modern parton showers and detector simulations for collider experiments.

Here, we focus on the latest upgrade described in detail in Ref.\ \cite{KKMCee:2023}, which includes a major rewriting in C++ as well as a significant improvement in the low-level MC generation process. The latest public version of KKMCee can be obtained from GitHub.\footnote{The public release of KKMCee can be obtained at \href{https://github.com/KrakowHEPSoft/KKMCee}{github.com/KrakowHEPSoft/KKMCee}.}  The new release is called KKMCee to distinguish it from the more recent hadronic branch  KKMChh~\cite{KKMChh:2016,KKMChh:2019} applying CEEX photonic corrections to quark scattering with leptonic final states.

KKMCee has been trans-coded from FORTRAN to C++. The class structure permits a more efficient and compact organization of the code implementing the complex spin amplitude calculations required for CEEX. However, the physics of CEEX is unaltered, as it remains the state of the art in precision photonic physics. 
CEEX naturally facilitates the inclusion of initial-final photon interference effects at all orders.
Narrow resonance effects important, for example, in initial-final interference near the $Z$ pole can be readily included. Also, CEEX makes a complete treatment of spin effects for both the beams and final unstable fermions (such as taus) possible.

The external DIZET and TAUOLA libraries remain in FORTRAN. DIZET was recently upgraded to version 6.45,~\cite{Arbuzov:2020coe} and is now called before using KKMCee to pretabulate electroweak form factors before a KKMCee run. This makes KKMCee more independent of the electroweak library than previously, and should facilitate an anticipated upgrade to an $O(\alpha^2)$ electroweak library.

ROOT~\cite{Brun:1997pa} now handles many basic services, including random number generation, Lorentz kinematics, histogramming, and graphics that were formerly handled by internal routines. ROOT also provides a useful persistence mechanism that permits more flexibility in the generation and analysis of events. It is possible to export the entire MC generator object to disk for analysis or to import this object from a disk file to resume a stopped run. 

The Monte Carlo algorithm underlying KKMC has been substantially upgraded using an adaptive MC program Foam,~\cite{Jadach:2002kn} which is capable of integrating an arbitrary multi-dimensional distribution supplied by the user. Foam uses a preliminary exploratory MC run to create a grid of rectangular (or simplicial) cells, denser in the regions where the distribution varies rapidly. Foam can efficiently find and account for peaks in the distribution automatically. Foam generates the ISR energy fraction, the beam energy spread parameters, the final fermion flavor, and the scattering angle of the final fermions with respect to the beams. 

\begin{figure}[ht]
\includegraphics[width=\textwidth]{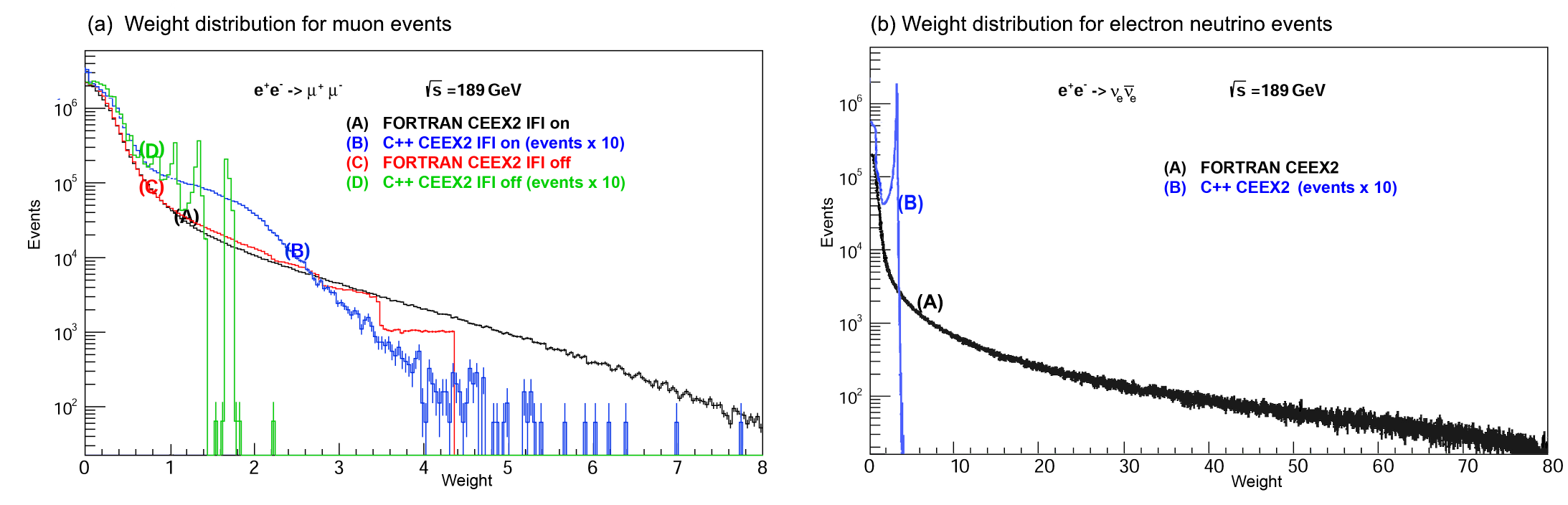}
\caption{Comparison of MC weights for the FORTRAN version and C++ version of KKMC for $\mu^+\mu^-$ events on the left and $\nu_e {\overline\nu}_e$ events on the right. Weights for the full CEEX2 matrix element are shown for the FORTRAN version in black (A) and the C++ version in blue (B). For the muon events, the weights without IFI are also shown in red (B) for the FORTRAN version and in green (D) for the C++ version. The KKMCee event counts are scaled by a factor of 10.} \label{fig:weights}
\end{figure}

Using Foam to generate the fermion scattering angle greatly improves the MC weight dispersion, leading to more efficient generation. Fig.\ \ref{fig:weights} (a) compares weight distributions for 2 million KKMCee 5.00 events to 18 million events from the FORTRAN version for $e^+ e^- \rightarrow \mu^+ \mu^- + n\gamma$ at a 189 GeV CM energy.\footnote{The KKMCee event counts are scaled by a factor of 10 to better compare the shapes of the distributions.} The weight distribution is about twice as compact with Foam. 
Fig.\ \ref{fig:weights} (b) compares similar weight distributions for $e^+ e^- \rightarrow \nu_e {\overline\nu}_e + n\gamma$. The effect of switching to Foam is most dramatic for $\nu_e$ generation, where a strong forward peak forms above 105 GeV due to $t$-channel $W$ exchange. Foam gives a factor of 20 improvement in the weight distribution in this case. 

\begin{figure}[hb]
\centering
\includegraphics[width=0.4\textwidth]{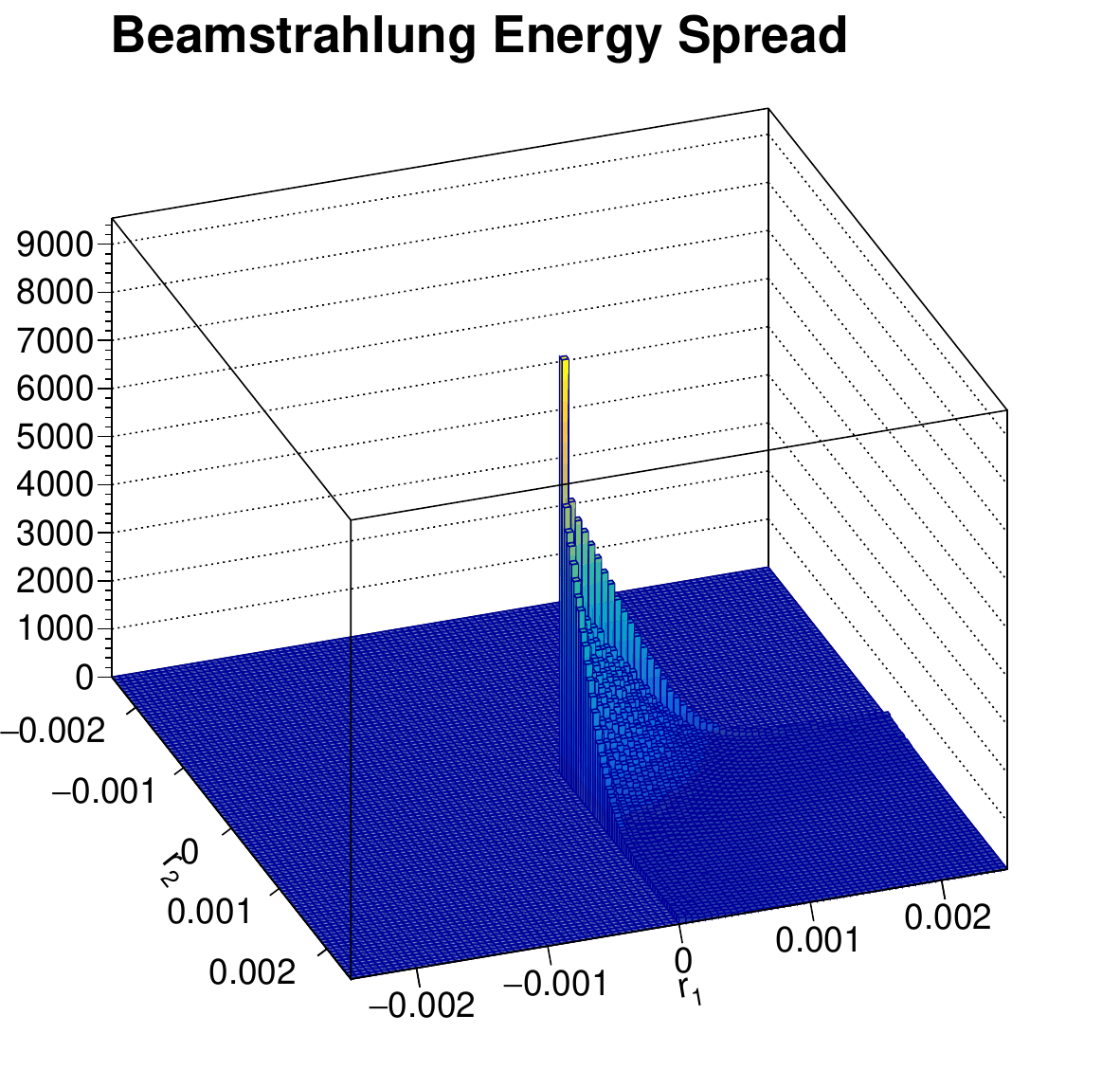}
\includegraphics[width=0.4\textwidth]{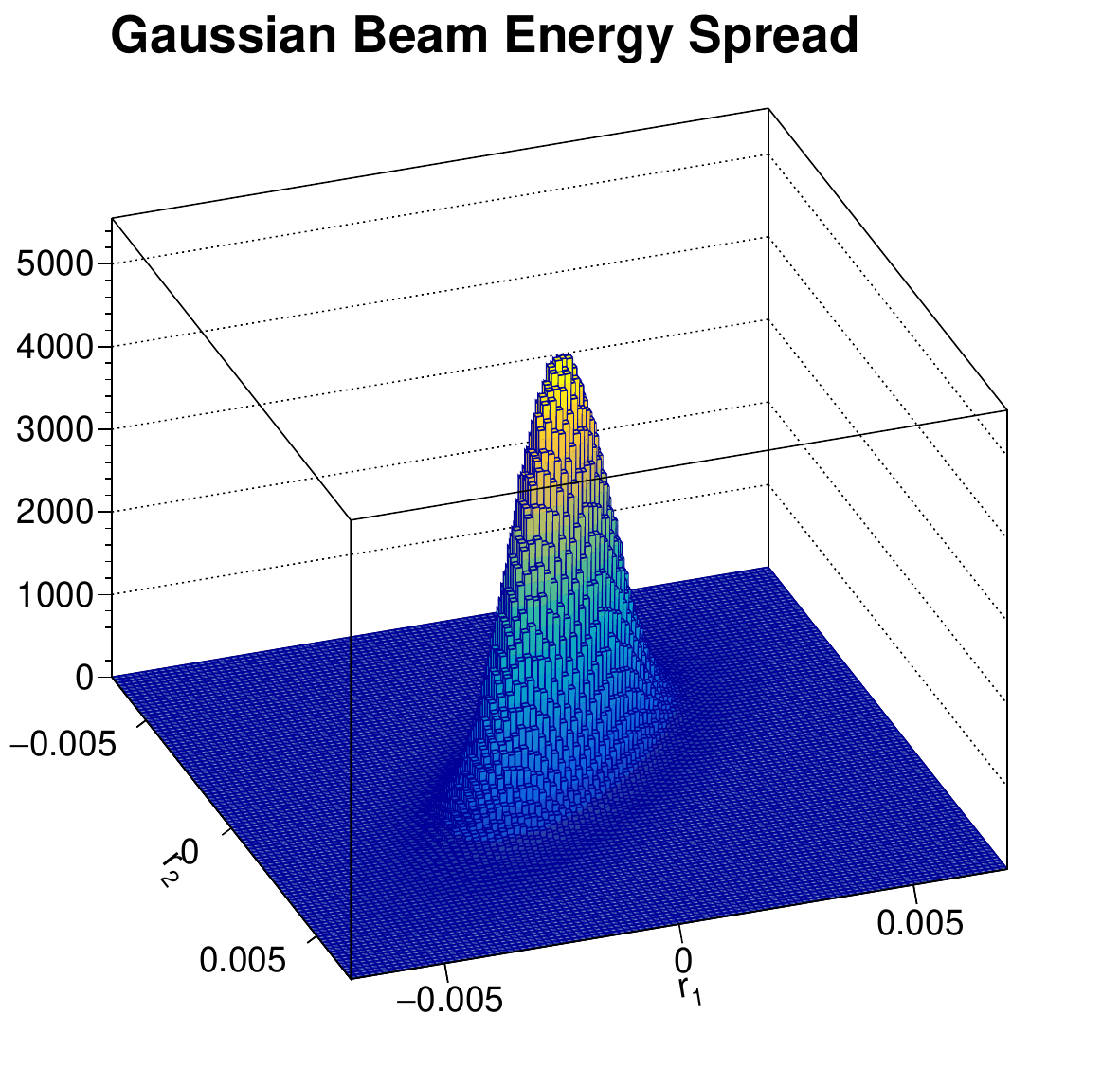}
\caption{Beamstrahlung and correlated Gaussian beam energy spread distributions of two beams in terms of $r_i=(E_i-\bar{E}_i)/\bar{E}_i$.}\label{fig:BES}
\end{figure}

KKMC previously supported beamstrahlung beam spread (BST) using variables $z_i = E_i/{\overline E}_i$ generated using a CIRCE1 parametrization~\cite{circe1} which has sharp peaks at $z_i = 1$. This BST distribution is now implemented in Foam. KKMCee now optionally includes correlated double-Gaussian beam energy spread (BES). The flexibility of Foam allows the user to provide a combined BES + BST distribution or other custom distribution. Fig.\ \ref{fig:BES} shows the BST and BES distributions in variables $r_i = 1-z_i$.

The auxiliary program KKsem~\cite{Jadach:2000ir} for comparing KKMC results with semi-analytical formulas is now replaced with the much more powerful tool KKeeFoam~\cite{Jadach:2018lwm} that includes resummed initial-final interference (IFI) in a semi-soft approximation in addition to the original initial-state radiation (ISR) and final-state radiation (FSR). All integrals over transverse photon degrees of freedom have been done analytically, as has the sum over photons to all orders, in a semi-soft approximation.   
KKeeFoam provides the invariant mass and angle of the final fermion pair and internally generates the total longitudinal momentum of the ISR and FSR photons. The addition of IFI, which was not available in KKsem, adds two new variables representing longitudinal momenta for interfering photons. KKeeFoam allows for independent switching of ISR, FSR and IFI. 

\begin{figure}[ht]
\centering
\includegraphics[width=0.85\textwidth]{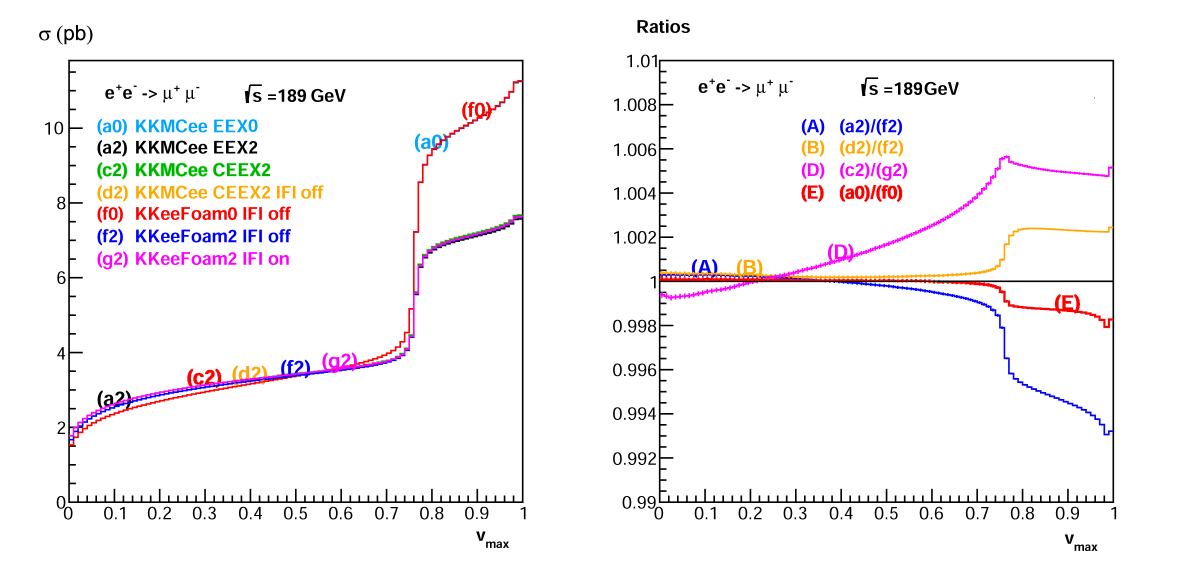}
\caption{
Sample cross-sections $\sigma(v_{\rm max})$ at $\sqrt{s} =189$ GeV as a function of the cutoff $v_{\rm max}$ on the total photonic energy fraction, comparing various types of QED matrix element for KKMCee
and KKeeFoam.
}
\label{fig:KKeeFoam}
\end{figure}

Fig.\ \ref{fig:KKeeFoam} shows the cross section $\sigma(v_{\rm max})$ for 7 billion events generated at a 189 GeV CM energy as a function of the cutoff on the total photon energy fraction $v=1-{s'}/s$, where $s$ and $s'$ and are the squared CM energy of the initial and final fermions. The numbers 0 and 2 denote the $O(\alpha^0)$ and $O(\alpha^2)$ matrix elements.

\begin{table}[ht]
\caption{Total Cross Section $\sigma(v_{\rm max})$ and charge asymmetry $A_{\rm FB}(v_{\rm max})$ at $\sqrt{s}=189$ GeV as a function of the maximum photonic energy fraction $v_{\rm max}$ for KKMC 4.13 (2000) and KKMCee 5.00 (2023).}
\label{tab1}\begin{center}\renewcommand{\arraystretch}{0.8}
\begin{tabular}{|c||c|c||c|c|}
\hline 
\raisebox{-1.5ex}[0cm][0cm]{$\mathbf{v_{max}}$} 
& \multicolumn{2}{c||}{$\mathbf{\mathbf{\sigma}(v_{max})}$\ \bf (pb)}
& \multicolumn{2}{c|}{$\mathbf{A_{\rm FB}(v_{\rm max})}$}\\\cline{2-5}
 & \small\bf KKMC 4.13 & \small\bf KKMCee 5.00 & \small\bf KKMC 4.13 & \small\bf KKMCee 5.00 \\
\hline 
\small 0.10 &\small $2.5967\pm 0.0027$ &\small $2.6015\pm 0.0001$ &\small $0.5922\pm 0.0012$ &\small $0.5931\pm 0.0001$ \\
\small 0.30 &\small $3.1190\pm 0.0029$ &\small $3.1236\pm 0.0001$ &\small $0.5856\pm 0.0011$ &\small $0.5864\pm 0.0001$ \\
\small 0.50 &\small $3.4203\pm 0.0029$ &\small $3.4250\pm 0.0001$ &\small $0.5863\pm 0.0010$ &\small $0.5970\pm 0.0000$ \\
\small 0.70 &\small $3.7596\pm 0.0030$ &\small $3.7641\pm 0.0001$ &\small $0.5947\pm 0.0009$  &\small $0.5953\pm 0.0000$ \\
\small 0.90 &\small $7.1780\pm 0.0030$ &\small $7.1849\pm 0.0001$ &\small $0.3170\pm 0.0005$  &\small $0.3174\pm 0.0000$ \\
\small 0.99 &\small $7.6542\pm 0.0029$ &\small $7.6596\pm 0.0001$ &\small $0.2912\pm 0.0004$  &\small $0.2917\pm 0.0000$ \\
\hline
\end{tabular}
\end{center}\end{table}

KKMCee 5.00 reproduces all of the previous KKMC benchmarks in Ref.~\cite{Jadach:2000ir}. Table 1 compares the KKMC 4.13 and KKMCee 5.00 results for the total cross section $\sigma(v_{\rm max})$ and charge asymmetry $A_{\rm FB}(v_{\rm max})$ as a function of the cutoff on the total photon energy fraction $v = 1-{s'}/s$ using the full CEEX $O(\alpha^2)$ matrix element at a 189 GeV CM energy. The 2023 results~\cite{KKMCee:2023} are calculated using 700 million unweighted events and the 2000 results are from Table III in Ref.~\cite{Jadach:2000ir}. 

The new release of KKMCee is intended to provide a flexible basis for future developments. Examples of anticipated upgrades include 
\begin{itemize}
\item{} adding CEEX $O(\alpha^3)$ at least to leading logarithm, which is already present in EEX,
\item{} forcing a visible photon at the generator level in neutrino pair channels,
\item{} automatic construction of the CEEX matrix element for porting to other processes such as $HZ$ production and decay,
\item{} replacing DIZET with a new $O(\alpha^2)$ electroweak library,
\item{} a more efficient algorithm for certain corners of phase space, such as two hard photons,
\item{} possibly integrating the Bhabha process into KKMCee.
\end{itemize}

\acknowledgments
This ICHEP contribution is dedicated to Stanisław Jadach (1947 -- 2023), an original creator of KKMC and leading force behind the upgrade to KKMCee.

KKMCee is supported in part by grant No.\ 2023/50/A/ST2/00224 of the National Science Center (NCN), Poland. 
S.A.\ Yost was supported in part by a grant from the Citadel Foundation. S.\ Jadach was supported in part
by the European Union’s Horizon 2020 Research and Innovation Program, Grant No.\ 951754 and the National Science Center, Poland, Grant No.\ 2019/34/E/ST2/00457. A. Si{\'o}dmok is supported in part by Priority Research Area Digiworld under the program ‘Ex-cellence Initiative – Research University’ at Jagiellonian University.

\end{document}